\documentclass{pasj01}

\Received{2018 June 19}
\Accepted{2018 July 30}
\Published{}

\begin{document}

\title{Constraints on accretion disk size in the massive type~1 quasar PG~2308+098 from optical continuum reverberation lags}
\author{Mitsuru \textsc{Kokubo} \altaffilmark{1}}%
\altaffiltext{1}{Astronomical Institute, Tohoku University, 6-3, Aramaki-Aza-Aoba, Aoba-ku, Sendai, Miyagi, 980-8578, Japan}
\email{mkokubo@astr.tohoku.ac.jp}


\maketitle

\begin{abstract}

Two years' worth of $u$-, $g$-, $r$-, $i$-, and $z$-band optical light curves were obtained for the massive type~1 quasar PG~2308+098 at $z=0.433$ using the 1.05-m Kiso Schmidt telescope/Kiso Wide Field Camera, and inter-band time lags of the light curves were measured.
Wavelength-dependent continuum reverberation lag signals of several tens of days relative to the $u$-band were detected at $g$-, $r$-, $i$-, and $z$-bands, where the longer wavelength bands showed larger lags.
From the wavelength-dependent lags, and assuming the standard disk temperature radial profile $T \propto R_{\rm disk}^{-3/4}$ and an X-ray/far-ultraviolet reprocessing picture, a constraint on the radius of the accretion disk responsible for the rest-frame 2500~\AA\ disk continuum emission was derived as $R_{\rm disk} = 9.46^{+0.29}_{-3.12}$~light-days.
The derived disk size is slightly ($1.2-1.8$ times) larger than the theoretical disk size of $R_{\rm disk} = 5.46$~light-days predicted from the black hole mass ($M_{\rm BH}$) and Eddington ratio estimates of PG~2308+098.
This result is roughly in accordance with previous studies of lower mass active galactic nuclei (AGNs), where measured disk sizes have been found to be larger than the standard disk model predictions by a factor of $\sim 3$; however, the disk size discrepancy is more modest in PG~2308+098.
By compiling literature values of the disk size constraints from continuum reverberation and gravitational microlensing observations for AGNs/quasars, we show that the $M_{\rm BH}$ dependence of $R_{\rm disk}$ is weaker than that expected from the standard disk model.
These observations suggest that the standard Shakura-Sunyaev accretion disk theory has limitations in describing AGN/quasar accretion disks.

\end{abstract}

\section{Introduction}

Intense radiation from active galactic nuclei (AGNs) at ultraviolet(UV)-optical wavelengths is produced in accretion disks formed around central supermassive black holes (SMBHs).
Luminous Seyfert galaxies and quasars are believed to harbor AGN accretion disks with sub-Eddington accretion rates, and the standard thin thermal accretion disk theory of \citet{sha73} is though to reasonably describe such sub-Eddington BH accretion disks.
However, debate has raged regarding the applicability of the standard accretion disk theory to AGNs, with claims of discrepancies between the observed spectral energy distribution of AGNs and the model's prediction (e.g., \cite{col02,dav07,kok14}; but see also \cite{kis08,cap15}).

Other than the emission spectrum, the geometry of accretion disks is another predictable property of BH accretion disks.
Based on standard accretion disk theory, the radial extent or size of a thin disk can be algebraically expressed as a function of the BH mass and mass accretion rate (e.g., \cite{poo07,mor10,fau16}); thus, the radial extent can be an essential observational probe to test the standard accretion disk theory in AGNs.
However, AGN accretion disks are physically compact, and the current largest optical telescopes are unable to directly spatially resolve such disks even in the nearest AGNs in the Local Universe.

Nevertheless, two indirect methods are available to measure accretion disk sizes in AGNs: one uses the microlensing phenomena of gravitationally-lensed quasars (e.g., \cite{mor10,bla11,jim12,cha16}) and the other uses inter-band lags of multi-band intrinsic flux variability (e.g., \cite{col99,ser05,cac07,ede15,fau16,bui17,fau18}).

The UV-optical microlensing effects observed in lensed high redshift quasars are caused by the lensing magnification of the quasar accretion disk emission by foreground stars in a lensing galaxy, which enables the accretion disk size to be inferred through modeling of the multi-epoch variability (e.g., \cite{mor10}) or single-epoch flux ratio anomalies (e.g., \cite{poo07,bla11}).

On the other hand, the intrinsic flux variability of normal (unlensed) Seyfert galaxies and quasars can also be used to measure disk sizes under the assumption of the `reprocessing' picture as the variability model.
In the reprocessing model, it is assumed that the intrinsic UV-optical continuum variability ubiquitously observed in AGNs is driven by intense, variable X-ray/far-UV (FUV) emission from a compact region in the vicinity of the SMBH (\cite{utt14,gar17} and references therein).
The variable X-ray/FUV emission irradiates the disk surface, altering the effective temperature of the disk and thus continuously altering the UV-optical flux.
It should be noted that the reprocessing diagram unsurprisingly predicts that the variation of the disk's effective temperature and corresponding disk luminosity at outer radii lags behind the variation at inner radii by light travel times.
Although the physical mechanism of the intrinsic UV-optical continuum variability continues to be debated, there is growing evidence that the multi-band light curves of AGNs show strong inter-band correlations with lags between the X-ray and UV-optical bands (e.g., \cite{ede15,fau16,bui17,fau18}) and the inter-band lags within the UV-optical bands (e.g., \cite{col99,ser05,cac07,jia17,mud17,mch17}) in a form consistent with $\tau \propto \lambda^{4/3}$, as expected from the reprocessing model based on the standard thin disk model.

Although recent intensive monitoring observations of the simultaneous X-ray, UV, and optical light curves for local Seyfert galaxies (e.g., NGC~4395, NGC~4593, NGC~5548, and NGC4151; see \cite{mch17} and references therein) have confirmed that the wavelength dependence of the continuum reverberation lags at the UV-optical bands is largely consistent with the standard thin disk model ($\tau \propto \lambda^{4/3}$), they have also revealed that the absolute disk sizes in these Seyfert galaxies inferred from the lags are generally larger by a factor of $\sim 3$ than those expected from the thin disk model, given the BH mass estimates for these objects (\cite{cac07,sha14,ede15,cac18}).
Interestingly, the microlensing constraints on the accretion disk sizes for lensed high-redshift quasars are also claimed to be larger than the thin disk predictions by a similar factor, on average (e.g., \cite{mor10,jim12,mos13,cha16}).
It has been proposed that local temperature fluctuations on the surface of the standard disk due to thermal instabilities can make the disk sizes measured via the microlensing technique unexpectedly large (\cite{dex11}), but such an inhomogeneous disk picture has difficulty in explaining the observed well-correlated UV-optical light curves (\cite{kok15}).
The discrepancies between the observed and model disk sizes, if confirmed, suggest that the standard thin disk model cannot provide an adequate explanation for actual AGN accretion disks.

One way to examine further the above-mentioned disk size problem is to probe disk sizes in AGNs with various BH masses and Eddington ratios by using the continuum reverberation lag method.
The X-ray-to-UV/optical simultaneous monitoring observations for the AGN continuum reverberation lag measurements carried out to date have targeted local Seyfert galaxies, which generally have lower BH masses and in many cases also have lower Eddington ratios compared to the lensed quasar sample (see e.g., \cite{ede15}).
It should be noted that continuum reverberation lags can be measured only with multi-band optical observations as inter-band lags at the rest-frame UV-optical bands; thus, optical time-domain wide-field surveys enable the measurement of continuum reverberation lags and, consequently, accretion disk sizes in field quasars with a higher BH mass and/or higher Eddington ratio compared to local Seyfert galaxies.
Recently, \citet{jia17} examined the optical inter-band disk continuum reverberation lags for 39 quasars using four years' worth of $g$-, $r$-, $i$-, and $z$-band light curves from the Pan-STARRS Medium Deep Survey and reported that the inter-band lags were generally $\sim 2-3$ times larger than those predicted by the thin disk model.
On the other hand, by examining the continuum lags from weekly sampled $g$-, $r$-, $i$-, and $z$-band light curves of 15 quasars with BH masses of $\sim 10^{8}-10^{9}$~$M_{\odot}$ in the Dark Energy Survey (DES) supernova fields, \citet{mud17} claimed that the disk sizes may be comparable to the standard thin disk model predictions.

In this work, we add a further quasar, PG~2308+098, to the sample of quasars whose continuum reverberation lags have been firmly detected.
We observed the $u$-, $g$-, $r$-, $i$-, and $z$-band light curves of PG~2308+098 at a redshift of $z=0.433$ (\cite{sch83,ves06}) for $\sim 800$ days, and detected inter-band continuum reverberation lags in this quasar.
To compare our disk size measurement result for PG~2308+098 with those of previous works, we adopt the same analysis method as that used by \citet{mud17}.
The BH mass of PG~2308+098 is $M_{\rm BH} = 10^{9.6 \pm 0.1} M_{\odot}$, which is more massive compared to the typical BH masses of the quasar samples of \citet{jia17} and \citet{mud17}.
Therefore, by adding PG~2308+098 to the sample of continuum reverberation mapped quasars, we can examine the BH mass vs. disk size relation over a much wider BH mass range than previous works.

Details of our monitoring observations for PG~2308+098 are described in Section~\ref{sec:observation}.
Section~\ref{sec:cont_rev_measure} outlines several analysis methods used to estimate the inter-band lags, and examines the consistency of the estimated lags derived from these analyses.
In Section~\ref{sec:disksize}, literature values of the disk size constraints from the continuum reverberation and microlensing observations for AGNs/quasars are compiled, and the disk size vs. black hole mass relation for AGNs/quasars is examined; we then demonstrate that the observed AGN/quasar disk sizes are in conflict with the standard accretion disk theory.
Implications for the accretion disk structures obtained from an examination of the disk size problem are discussed in Section~\ref{sec:implication}.
Section~\ref{sec:conclusion} provides a summary and conclusions.

\section{Observations}
\label{sec:observation}

\begin{figure}[tbp]
\center{
\includegraphics[clip, width=3.1in]{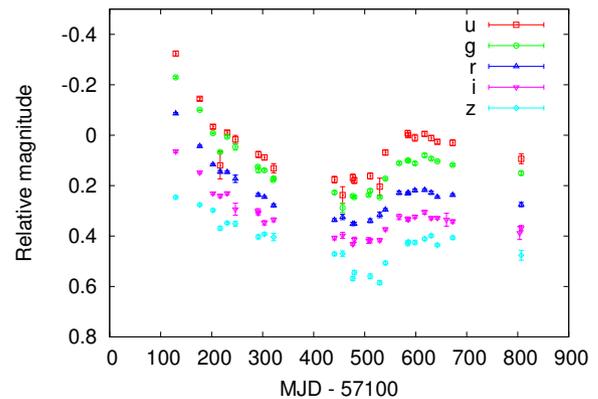}
}
 \caption{$u$-, $g$-, $r$-, $i$-, and $z$-band light curves of PG~2308+098 obtained with the 1.05-m Kiso Schmidt Telescope/Kiso Wide Field Camera. 
The magnitudes are arbitrarily shifted, for clarity.}
 \label{fig:lc}
\end{figure}

\subsection{Observational properties of PG~2308+098}
\label{object}

PG~2308+098, also known as 4C~09.72, is a bright ($V\sim 16$~mag) type~1 quasar at a redshift of $z=0.433$.
A BH mass estimate for PG~2308+098 is given by \citet{ves06} as $\log (M_{BH}/M_{\odot})=9.6 \pm 0.1$, which is based on a broad H$\beta$ line width taken from \citet{bor92} and a monochromatic luminosity at 5100~\AA\ of $\log (\lambda L_{\lambda}~{\rm [ergs/s]}) = 45.8 \pm 0.1$ taken from \citet{sch83}.
Using a bolometric correction factor of ${\rm BC}_{5100} = 9.26$ (\cite{ric06}), the bolometric luminosity and Eddington ratio of PG~2308+098 are estimated to be $\log (L_{\rm bol} [\rm ergs/s]) = 46.8 \pm 0.1$ and $\log (L_{\rm bol}/L_{\rm Edd}) = -0.9 \pm 0.1$, respectively.

\citet{kis08} showed that the optical-near infrared polarized flux spectrum of PG~2308+098 is consistent with that of a power-law shape, $f_{\nu} \propto \nu^{\alpha_{\nu}}$, with a power-law index of $\alpha_{\nu} \simeq 1/3$.
The power-law index of the optical spectrum of the AGN accretion disk is related to the power-law index of the disk temperature radial profile:
\begin{equation}
T \propto R_{\rm disk}^{-1/\beta}
\label{eqn:temp_radius}
\end{equation}
where $\alpha_{\nu} = 3-2\beta$, on the assumption of multi-temperature black body disks (e.g., \cite{kat08}), and the standard thin disk model presented by \citet{sha73} predicts $\beta=4/3$ ($\alpha_{\nu}=1/3$).
Therefore, the spectral index of the polarized flux spectrum of PG~2308+098 suggests that the disk temperature profile of PG~2308+098 is consistent with the standard disk model prediction.
\citet{kok16} carried out $u$-, $g$-, $r$-, $i$-, and $z$-band monitoring observations for PG~2308+098 and showed that although the detailed spectral shape at rest-frame UV wavelengths differs between the polarized and variable component spectra (see \cite{kok16,kok17} for details), the optical variable component spectrum is also roughly consistent with a power-law spectral shape of $\alpha_{\nu} \simeq 1/3$.
We note that power-law variable component spectra with $\alpha_{\nu} \simeq 1/3$ are generally observed in type~1 AGNs/quasars (see e.g., \cite{kok14,mac16,bui17}).
These observations suggest that the accretion disk in PG~2308+098 might be described by the standard thin disk model.

\subsection{Five-band optical monitoring observations with the 1.05-m Kiso Schmidt Telescope/KWFC}
\label{sec:obs}

\begin{table}[tbp]
\tbl{Kiso/KWFC $u$-, $g$-, $r$-, $i$-, and $z$-band light curves of PG~2308+098.}{%
\begin{tabular}{cccc}  
\hline\noalign{\vskip3pt} 
MJD & Magnitude & Error in magnitude & Band \\  [2pt] 
\hline\noalign{\vskip3pt} 
57229.588 &   -0.324 &    0.009 & $u$ \\
57276.657 &   -0.144 &    0.007 & $u$ \\
57302.468 &   -0.034 &    0.008 & $u$ \\
$\cdots$   &$\cdots$ &$\cdots$ &  $\cdots$\\ [2pt]  
\hline\noalign{\vskip3pt} 
\end{tabular}}\label{lightucrve_table}
\begin{tabnote}
\hangindent6pt\noindent
\hbox to6pt{\footnotemark[$*$]\hss}\unskip%
 The light curve in each band is shifted so that the average magnitude is zero. The complete listing of this table is available in the online edition as Supporting Information.
\end{tabnote}
\end{table}

The $u$-, $g$-, $r$-, $i$-, and $z$-band optical light curves of PG~2308+098 were obtained during the period April 2015 to June 2017 using the Kiso Wide Field Camera (KWFC) mounted on the 1.05-m Kiso Schmidt Telescope at the Kiso Observatory, Japan \citep{sak12}\footnote{Longitude = 137.625406${}^{\circ}$E,  Latitude = 35.797236${}^{\circ}$N, and Altitude = 1130~m.}.
Part of the data obtained during the period $2015-2016$ has already been presented in \citet{kok16}; here, we re-analyze the full dataset, which includes data obtained after the publication of \citet{kok16}.
It should be noted that strong broad emission lines, such as Mg~II, H$\beta$, and H$\alpha$, are not strongly contaminated in the five bands at the redshift of PG~2308+098 (see e.g., \cite{jia17}), although the hydrogen Balmer continuum may possibly be a non-negligible contaminant to the $g$-band light curve (\cite{kok14,ede15,cac18}).
Possible contamination from the Balmer continuum will be investigated in Section~\ref{sec:comparison}.

The KWFC is a mosaic charge-coupled device (CCD) camera composed of four MIT and four SITe 2k$\times$4k CCD chips, and its full field-of-view is approximately $2^{\circ} \times 2^{\circ}$ (see \cite{sak12} for details).
During our observations, the KWFC was operated in a fast readout mode, where only the four MIT chips are used and $2\times2$~pixel on-chip binning is applied.
In this 28-s fast-readout mode, the pixel scale is 1.89 arcsec~pixel${}^{-1}$ and the field of view is limited to $1^{\circ} \times 2^{\circ}$.
The four MIT chips are arranged in a 2$\times$2 mosaic roughly aligned to the North-East on-sky direction; South-West, South-East, North-West, and North-East are referred to as chip\#0, chip\#1, chip\#2, and chip\#3, respectively.
Each of the chips has its own particular characteristics.
Chip\#0, chip\#1, chip\#2, and chip\#3 have gain values of 2.05, 2.30, 2.10, and 2.30 electron~ADU${}^{-1}$ and readout noises of 6.8,  6.9, 11.6, and 15.0 electrons, respectively.
\citet{kok16} analyzed only chip\#3 images because the pointing of the telescope was always set to introduce the target into chip\#3.
However, due to telescope pointing errors, the target was sometimes introduced into the other chips instead, especially into chip\#1.
Therefore, in this work we also analyzed the images of the other chips to ensure that all the photometry data obtained during the course of our observation runs were included.

During a single visit for each target, the five-band images were obtained quasi-simultaneously in the order $g$, $i$, $r$, $u$, and $z$.
For each of the five bands, four images with slight spatial offsets (1~arcmin four-point dithering) were obtained.
The exposure time for each image was 30~s for every filter during April-June 2015, and 60~s for the $g$-, $r$-, and $i$-bands and 120~s for the $u$- and $z$-bands during July 2015-June 2017.

Data reduction was performed by using custom-made python and IRAF/PyRAF scripts\footnote{IRAF is distributed by the National Optical Astronomy Observatory, which is operated by the Association of Universities for Research in Astronomy (AURA) under a cooperative agreement with the National Science Foundation.}, which were slightly modified in parts from the scripts used in \citet{kok16}.
Each of the CCD chips had a dual amplifier readout, and we treated the two readout areas (corresponding to the two amplifiers) separately, considering slightly different gain factors associated with the different amplifiers.
Overscan and bias subtraction, and pixel flat (= dome flat) and illumination flat (= sky flat) corrections were applied in the same way as in \citet{kok16}.
Additionally, the L.A.Cosmic algorithm (implemented in an IRAF task {\tt lacos\_im}; \cite{dok01}) was used to create mask files for pixels affected by cosmic-rays.
Sky coordinates of the images were calibrated by using the {\tt Astrometry.net} code and bundled {\tt image2xy} program \citep{lan10} and index files built from the Two Micron All Sky Survey (2MASS) catalog \citep{skr06}.

Aperture photometry was applied by using {\tt SExtractor} \citep{ber96}.
{\tt SExtractor} was applied twice for each image: in the first run, a seeing full-width-at-half-maximum (FWHM) size was estimated as a median of the FWHM values listed in SExtractor's output catalog\footnote{The median and the 16 and 84~\% percentiles of the seeing FWHMs at the $u$, $g$, $r$, $i$, and $z$ bands evaluated from all the images are $4.8^{+1.1}_{-0.8}$, $4.5^{+1.0}_{-0.8}$, $4.2^{+1.5}_{-0.9}$, $4.0^{+1.4}_{-0.7}$, $3.8^{+0.8}_{-0.6}$ arcsec, respectively.}; in the second run, the circular aperture diameter was adjusted to twice the seeing FWHM size to maximize the signal-to-noise ratio of the photometric measurements and minimize the effects of systematic centering errors (e.g., \cite{mig99}).
Photometric measurements affected by cosmic rays identified by {\tt lacos\_im} were excluded from the following analysis.

Zero-point offsets of the images were calibrated for each filter by performing inhomogeneous ensemble photometry (\cite{honey92}), using the {\tt Ensemble} software developed by Michael W. Richmond (e.g., \cite{vaz15}) \footnote{http://spiff.rit.edu/ensemble/; accessed December 12, 2017.}.
As comparison stars, field stars within a distance from PG~2308+098 appearing in each frame (8, 5, 3, 3, and 3 arcminutes for $u$, $g$, $r$, $i$, and $z$-band) were used to reduce the possible effect of flat-field errors.
The same set of comparison stars did not always appear on the images due to the uncertainties of the telescope pointings, and only stars appearing in at least 50~\% of the frames were used.
Inhomogeneous ensemble photometry is a procedure used to find zero-point offsets for each image to minimize the variance of the light curves of the comparison stars.
The absolute magnitude scale is arbitrary, and {\tt Ensemble} sets the magnitude of the brightest star to be zero.

The time stamp of each image was evaluated at mid-exposure, where the reference time frame and the time standard were set to the Barycentric Julian Date (BJD) and Barycentric Dynamical Time (TDB), respectively, using {\tt barycorrpy} Version 0.2 \citep{eas10,kan18}\footnote{https://github.com/shbhuk/barycorrpy}.
The time stamp is expressed as the Modified Julian Date (MJD), defined as MJD = BJD${}_{\rm TDB} - 2400000.5$.
Multiple exposures obtained during a single visit were binned by taking a weighted average after a $5\sigma$ clipping.

Figure~\ref{fig:lc} and Table~\ref{lightucrve_table} present the five-band light curves of PG~2308+098.
The numbers of observations of the light curves analysed in this work are 24, 28, 25, 29, and 24 for $u$-, $g$-, $r$-, $i$-, and $z$-band, respectively, and the median cadence is $\sim 16$~days.
The light curves of PG~2308+098 show a smooth decline from July 2016 to August 2016 (MJD $\sim$ $57,200-57,600$), and then quickly increase until November 2016 (MJD $\sim$ $57,600-57,700$).
These concave-shaped light curves enable us to confidently constrain the inter-band lags.

\section{Continuum reverberation lag measurements}
\label{sec:cont_rev_measure}

\subsection{JAVELIN thin disk reprocessing model}
\label{javelin_thin_disk}

The stochastic behavior of the UV-optical light curves of AGNs/quasars is known to be well modeled by first-order autoregressive Gaussian processes  or a damped random walk (DRW) model (\cite{mac10,sub17} and references therein).
A DRW time series is defined by an exponential covariance function in the form $\sigma_{\rm DRW}^2 \exp (-\Delta t/\tau_{\rm DRW})$, where $\Delta t$ is the time separating two observations and $\sigma_{\rm DRW}$ and $\tau_{\rm DRW}$ are the DRW parameters (\cite{zu13}).
DRW modeling of AGN light curves is known to be a powerful tool that interpolates sparsely sampled AGN/quasar light curves, thus enabling the reliable measurement of lags between two light curves (e.g., \cite{zu11,fau16}).

JAVELIN\footnote{https://bitbucket.org/nye17/javelin} (Just Another Vehicle for Estimating Lags In Nuclei) is a publicly available python code for estimating reverberation lags for AGN light curves\footnote{JAVELIN internally uses a Markov Chain Monte Carlo (MCMC) Ensemble sampler {\tt emcee} (\cite{for13}).}.
JAVELIN was originally designed to derive the posterior distribution (through MCMC calculations) of the reverberation lags between the continuum and the broad emission line light curves of AGNs/quasars, assuming the DRW model for the continuum light curve and top-hat transfer functions (see \cite{zu11} for details).
Currently, JAVELIN is widely used not only for broad emission line reverberation lag measurements (e.g., \cite{zu11,zu13}), but also for inter-band continuum reverberation lag measurements (e.g., \cite{sha14,fau16,starkey17,jia17,mud17,fau18}).

\citet{mud17} added a new function to JAVELIN (JAVELIN thin disk model; since version 0.33), which simultaneously fits the disk reprocessing model to multi-band continuum light curves on the assumption of the functional form of:
\begin{equation}
\tau \propto \lambda^{\beta},
\label{eqn:tau_lambda}
\end{equation}
where $\tau$ is the wavelength-dependent continuum reverberation lag relative to the variations of the driving X-ray/FUV emission (see \cite{mud17} for details).
Equation~\ref{eqn:tau_lambda}, combined with Wien's displacement law ($T \propto \lambda^{-1}$), represents the light-crossing time of the disk with a temperature radial profile described by Equation~\ref{eqn:temp_radius}, while $\beta = 4/3$ corresponds to the standard thin disk model.
\citet{mud17} used this new JAVELIN function to derive the continuum lags of the response light curve at $\lambda_{\rm rest} = 2500$~\AA\ relative to the driving light curve (at $\lambda \rightarrow 0$~\AA) for the 15 DES quasars, by fixing $\beta$ at the thin disk value of $\beta=4/3$.
Following \citet{mud17}, we applied the JAVELIN disk reprocessing model to the five band optical light curves of PG~2308+098.

\begin{figure}[tbp]
\center{
\includegraphics[clip, width=3.2in]{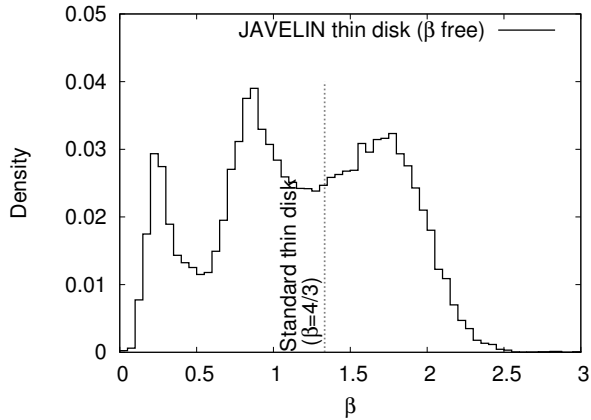}
}
 \caption{The posterior distribution of $\beta$ for PG~2308+098, constrained from the JAVELIN thin disk model analysis without fixing $\beta$. $\beta=4/3$ corresponds to the standard thin disk model.}
 \label{fig:beta_not_fixed}
\end{figure}

First, we tried to constrain $\beta$ by fitting the five band light curves of PG~2308+098 to the JAVELIN thin disk model without any constraint on $\beta$.
We adopted the $u$-band light curve as the driving light curve, and the other four band light curves were taken to be the response light curves.
There are six model parameters in the JAVELIN thin disk model: the power-law index $\beta$, the DRW parameters for the $u$-band (i.e., $\sigma_{\rm DRW}$ and $\tau_{\rm DRW}$), the lag of the $u$-band light curve relative to the driving light curve, and the scaling factor and kernel width of the transfer function for each of the $g$-, $r$-, $i$-, and $z$-bands.
A hard boundary from $100$ to $300$~days for the prior distribution of the rest-frame DRW timescale $\tau_{\rm DRW}$ was imposed, as suggested by \citet{mud17}, considering an observational fact that the typical value of $\tau_{\rm DRW}$ is $\sim$ 200 days for luminous quasars (e.g., \cite{mac10}).
The allowed observed-frame lags of the $g$-, $r$-, $i$-, and $z$-band light curves were restricted to a range from $-50$ to $50$~days.
MCMC sampling parameters {\tt nwalker}, {\tt nchain}, and {\tt nburn} (numbers of random walkers, sampling iterations, and burn-in iterations for each walker, respectively) were set to 1000, 1000, and 100 (see \cite{for13} for details).
Figure~\ref{fig:beta_not_fixed} shows the posterior distribution of $\beta$ constrained from the five band light curves of PG~2308+098 with JAVELIN thin disk model fitting.
Although the distribution is not inconsistent with the thin disk model value of $\beta=4/3$, the distribution widely spreads from $\beta \sim$ 0.2 to 2 with no clear single peak, so it is impossible to derive a point estimate for $\beta$ from this distribution.
This analysis indicates that it is difficult to constrain $\beta$ from multi-band light curves at rest-frame UV-optical wavelengths alone.
Therefore, following \citet{mud17}, in the remainder of this study we derive constraints on the continuum reverberation lag (and equivalently the disk size) by fixing $\beta = 4/3$.
In the case of PG~2308+098, the assumption of $\beta = 4/3$ is supported by the fact that the power-law spectral shape of the optical polarized flux spectrum and the variable component spectrum is consistent with the thin disk model prediction of $\alpha_{\nu} = 1/3$, as noted in Section~\ref{object}.
The possibility of $\beta \neq 4/3$ will be revisited in Section~\ref{sec:implication}.

\begin{figure}[tbp]
\center{
\includegraphics[clip, width=3.2in]{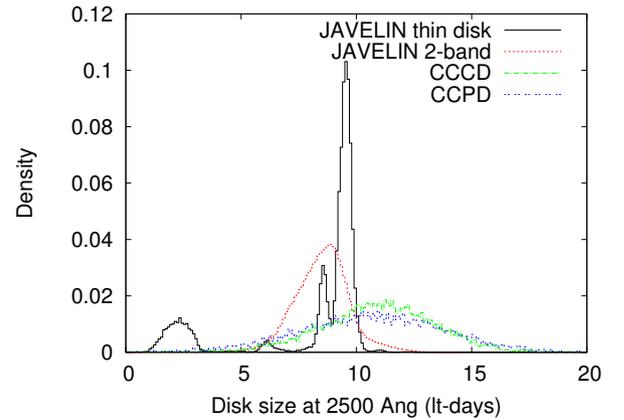}
}
 \caption{The probability distributions of the disk size at $\lambda_{\rm rest} = 2500$~\AA, $R_{\rm disk} = c\tau_{\rm rest, 2500}$, on the assumption that the multi-band lag is proportional to $\lambda^{4/3}$ ($\beta=4/3$). The distribution of the JAVELIN two-band analysis is calculated from the distributions of the lags of the four bands relative to the $u$-band (Figure~\ref{fig:javelin_posterior}), and those of cross-correlation centroid distributions (CCCDs) and cross-correlation peak distributions (CCPDs) are calculated from the CCCDs and CCPDs of the four bands (Figure~\ref{fig:ccfs}), respectively.}
 \label{fig:all_lag}
\end{figure}

Then, we derived the posterior distribution of the continuum reverberation lag at $\lambda_{\rm rest} = 2500$~\AA\ by fixing $\beta = 4/3$ in the JAVELIN thin disk model.
From the MCMC draws of the observed-frame lag of the $u$-band light curve relative to the driving light curve (referred to as $\tau_{{\rm obs, u}}$), we derived a probability distribution of the continuum lag at $\lambda_{\rm rest}=2500$~\AA\ (referred to as $\tau_{{\rm rest,~2500}}$), defined as:
\begin{equation}
\tau_{{\rm rest,~2500}} = \tau_{{\rm rest, u}} \left(\frac{2500~{\rm \AA}}{\lambda_{u, {\rm rest}}}\right)^{4/3} = \frac{\tau_{{\rm obs, u}}}{1+z}\left(\frac{2500~{\rm \AA}}{\lambda_{u, {\rm rest}}}\right)^{4/3},
\label{disk_at_2500}
\end{equation}
where $\lambda_{u, {\rm rest}} = 3551~{\rm \AA}/(1+z)$.
Figure~\ref{fig:all_lag} presents the probability distribution of $\tau_{{\rm rest,~2500}}$ derived from the above analysis.
It shows a strong peak at $\tau_{{\rm rest,~2500}} \sim 9.5$~days, although there is a sub peak at around $2.5$~days.
The median and $\pm 1\sigma$ (68.2~\% range) around the median are evaluated to be $\tau_{{\rm rest,~2500}} = 9.46^{+0.29}_{-3.12}$~days, and the 95th percentile upper limit is 9.91~days (Table~\ref{disksize}).
On the assumption of the reprocessing model, $\tau_{{\rm rest,~2500}}$ directly corresponds to the accretion disk size responsible for the disk continuum emission at $\lambda_{\rm rest} = 2500$~\AA\ of $R_{\rm disk} = c\tau_{{\rm rest,~2500}} = 9.46^{+0.29}_{-3.12}$~light-days.

\subsection{JAVELIN for two-band light curve pairs}
\label{javelin_twoband}

\begin{figure}[tbp]
\center{
\includegraphics[clip, width=3.2in]{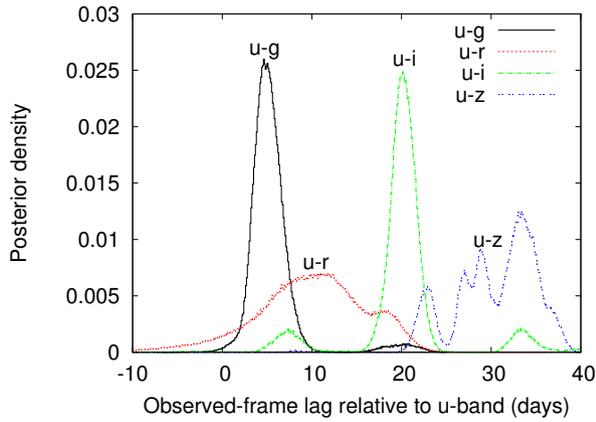}
}
 \caption{The posterior distributions of the observed-frame lag relative to the $u$-band light curve, evaluated for $u-g$, $u-r$, $u-i$, and $u-z$ band pair light curves using JAVELIN two-band analysis (Section~\ref{javelin_twoband}).}
 \label{fig:javelin_posterior}
\end{figure}

JAVELIN was originally designed to derive a reverberation lag between a pair of two-band light curves.
Here, we use JAVELIN's original function to constrain the inter-band lags of the disk continuum of the pairs of two-band light curves of PG~2308+098 to examine the consistency between the result from the JAVELIN thin disk analysis and that derived from the normal JAVELIN two-band analysis.

We again took the shortest wavelength band, i.e., the $u$-band, as the driving light curve of the disk reprocessing.
This means that the other four band light curves are assumed to be shifted, scaled, and smoothed versions of the $u$-band light curve.
First, the $u$-band light curve alone was fitted using the Cont\_Model module in JAVELIN to derive constraints on the DRW parameters.
Again, the rest-frame DRW timescale $\tau_{\rm DRW}$ was restricted to 100-300 days when constraining the continuum light curve model, as assumed by \citet{mud17}.
Then, each of the two-band pairs of light curves, $u-g$, $u-r$, $u-i$, and $u-z$, was fitted using the Rmap\_Model module in JAVELIN to derive the posterior distribution of the observed-frame lag between the two light curves.
Again, the observed-frame lags of the four band light curves were restricted to  a range from $-50$ to $50$~days.

Figure~\ref{fig:javelin_posterior} presents the posterior distributions of the observed-frame lag relative to the $u$-band light curve, evaluated for the $u-g$, $u-r$, $u-i$, and $u-z$ band pair of light curves.
Although the distributions are not single-peaked and have wide widths, a trend can be identified whereby the lags become larger at longer wavelength bands.
On the assumption of the wavelength dependence of the lag as $\tau \propto \lambda^{4/3}$ (i.e., $\beta = 4/3$), the lag relative to the $u$-band light curve as a function of the rest-frame wavelength can be expressed as:
\begin{eqnarray}
&&\tau_{{\rm rest}} (\lambda_{{\rm rest}}) - \tau_{{\rm rest}} (\lambda_{u, {\rm rest}}) \nonumber\\
&=& \tau_{{\rm rest,~2500}} \left(\frac{\lambda_{u, {\rm rest}}}{2500~{\rm \AA}}\right)^{4/3} \left [ \left(\frac{\lambda_{{\rm rest}}}{\lambda_{u, {\rm rest}}}\right)^{4/3} - 1 \right],
\label{lag_relative_to_u}
\end{eqnarray}
where $\lambda_{u, {\rm rest}} = 3551~{\rm \AA}/(1+z) = 2478$~\AA, and $\lambda_{\rm rest}$ takes values of (4686~\AA, 6166~\AA, 7480~\AA, 8932~\AA)$/(1+z)$.
$\tau_{{\rm rest,~2500}}$, which appears in Equation~\ref{lag_relative_to_u}, is the same as in Equation~\ref{disk_at_2500}.
By fitting Equation~\ref{lag_relative_to_u} to each set of the four MCMC draws of the $u-g$, $u-r$, $u-i$, and $u-z$ two-band lags produced by the JAVELIN two-band analyses, we derive a probability distribution of $\tau_{{\rm rest,~2500}}$, which is shown in Figure~\ref{fig:all_lag}.
The median and $\pm 1\sigma$ values are evaluated to be $\tau_{{\rm rest,~2500}} = 8.57^{+0.99}_{-1.20}$~days, and the 95th percentile upper limit is $10.42$~days (Table~\ref{disksize}).
The value of $\tau_{{\rm rest,~2500}}$ estimated from the JAVELIN two-band pairs is consistent with that estimated from the JAVELIN thin disk model (Table~\ref{disksize}), although the probability distributions shown in Figure~\ref{fig:all_lag} differ slightly from each other.

\subsection{Cross-correlation analysis}
\label{sec:ccf}

\begin{figure}[tbp]
\center{
\includegraphics[clip, width=3.2in]{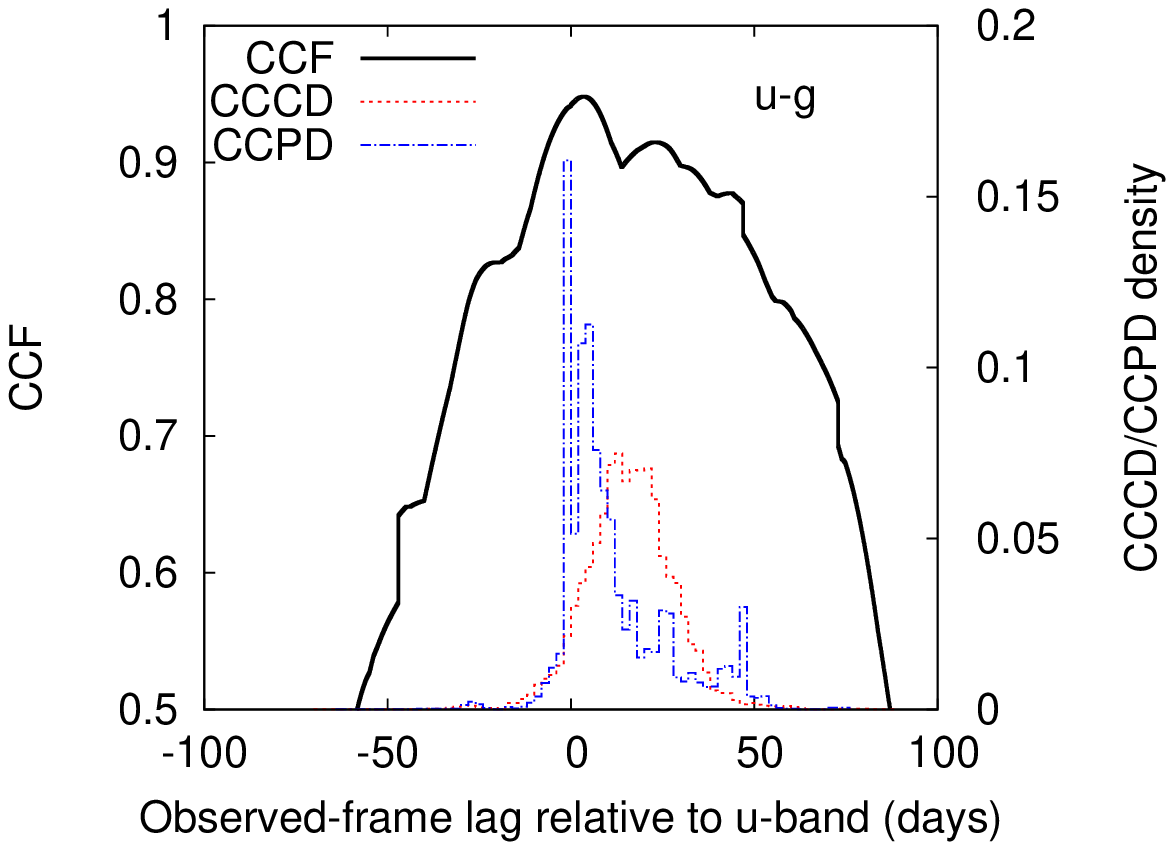}
\includegraphics[clip, width=3.2in]{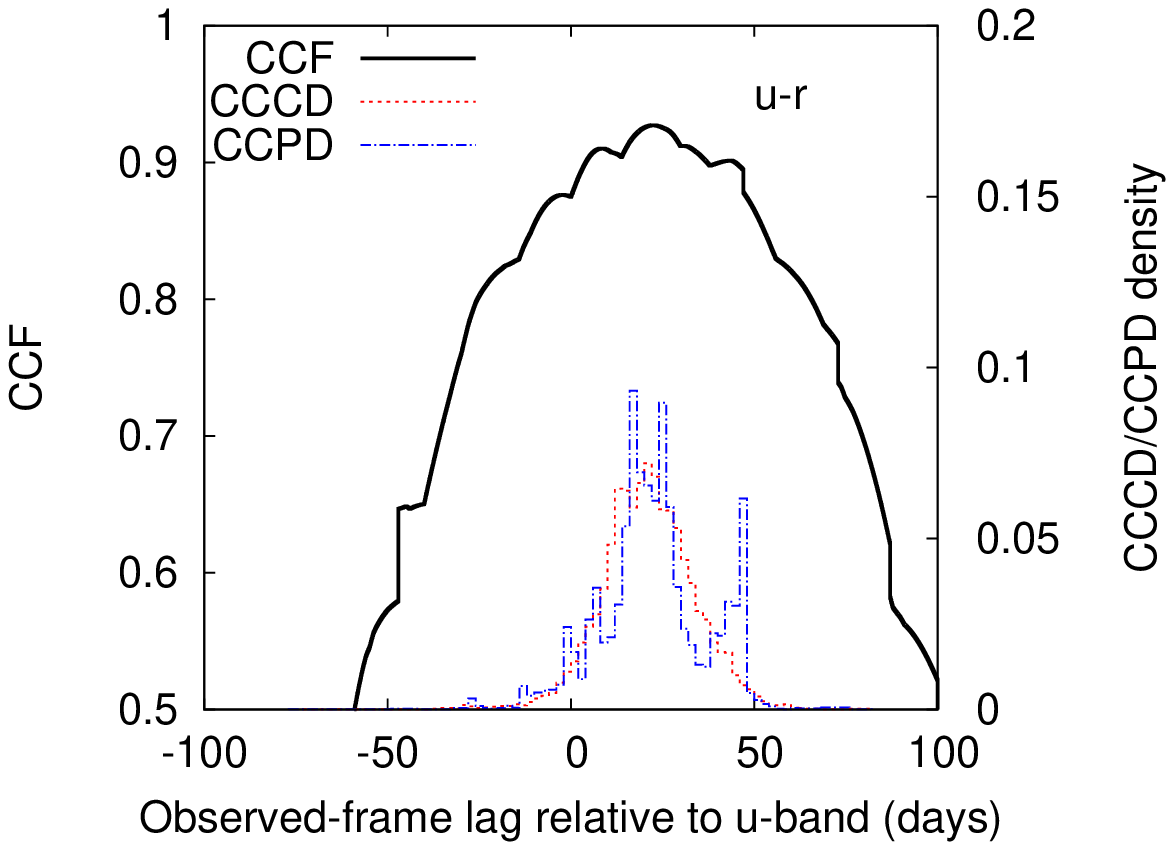}
\includegraphics[clip, width=3.2in]{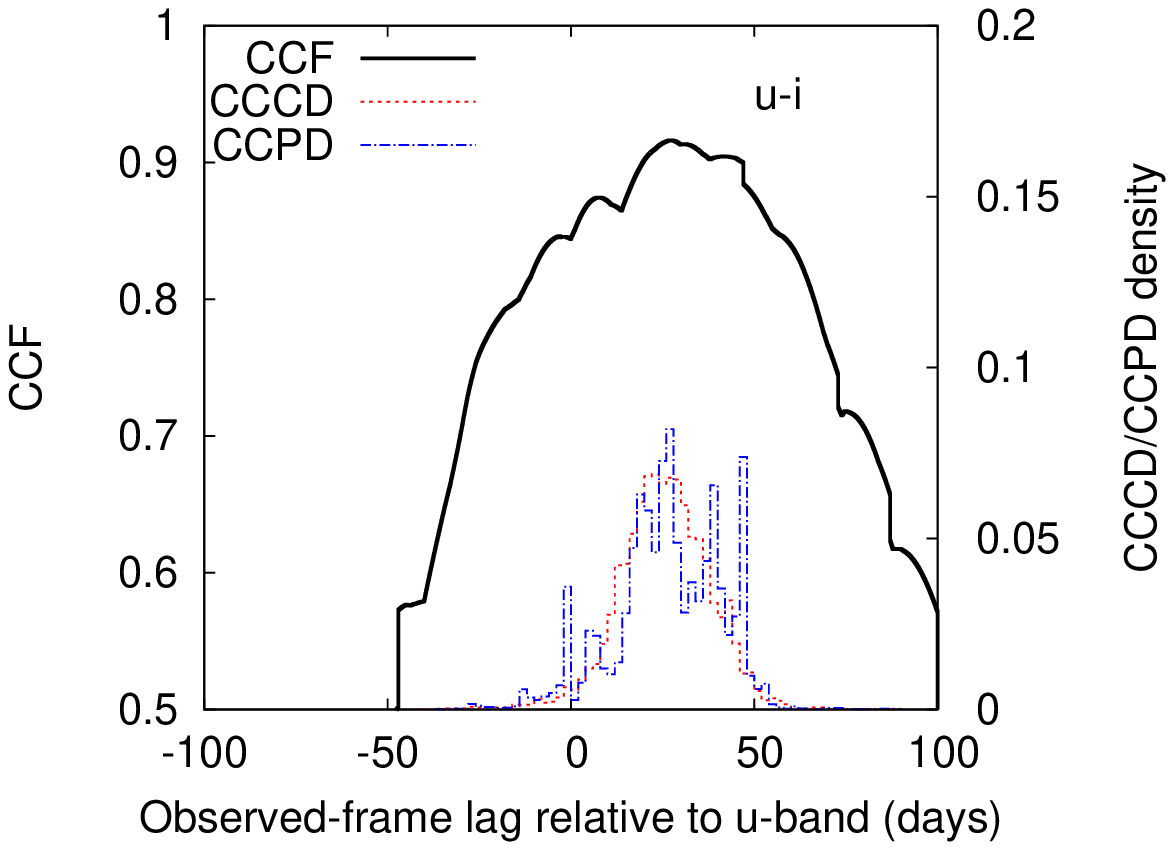}
\includegraphics[clip, width=3.2in]{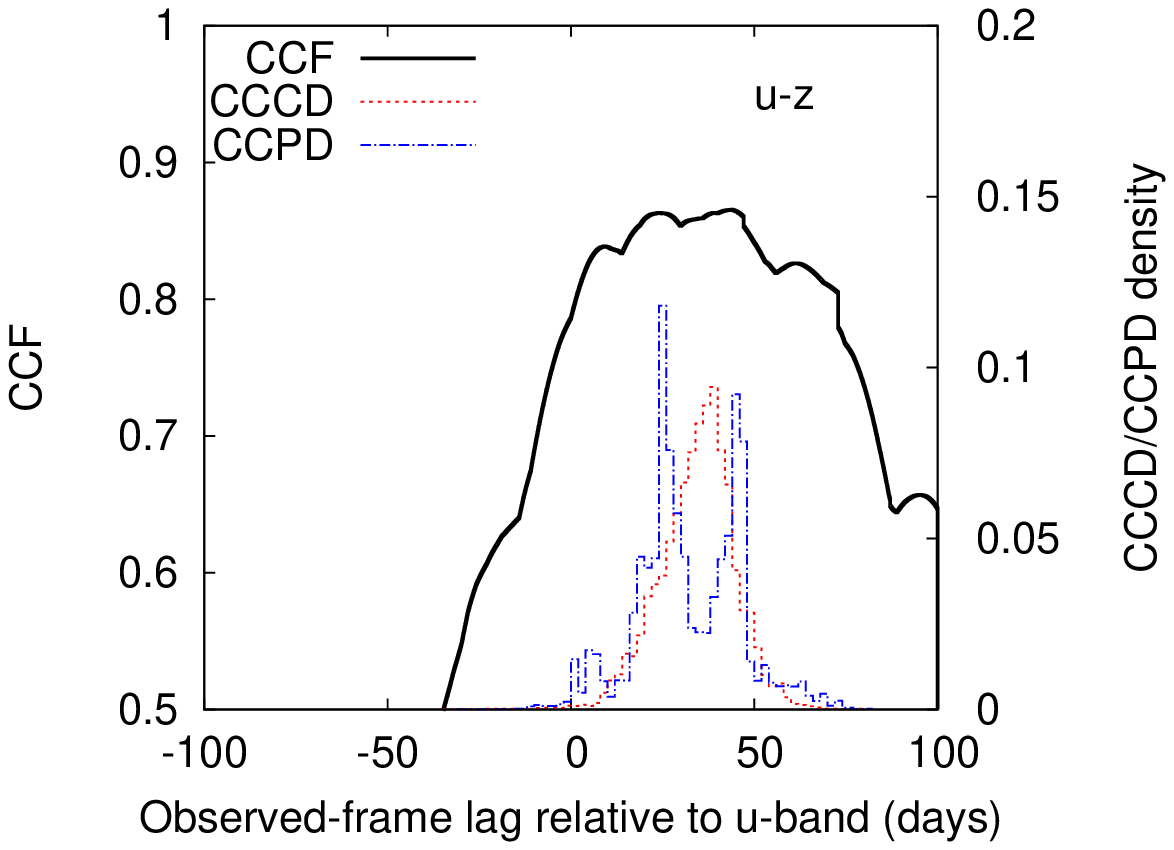}
}
 \caption{The cross-correlation functions (CCFs), CCCDs, and CCPDs of the $u-g$, $u-r$, $u-i$, and $u-z$ band pair light curves of PG~2308+098. A positive lag is defined as a positive observed-frame lag relative to the $u$-band light curve.}
 \label{fig:ccfs}
\end{figure}

Cross-correlation analysis is a classical method used to determine a temporal shift between two light curves (\cite{pet98,wel99}).
A temporal shift between two light curves can be evaluated by cross-correlating them and finding the lag where the correlation is maximized.
Here we explore whether the model-independent cross-correlation method results in lag values consistent with those derived from the DRW-based JAVELIN analyses.

Cross-correlation functions (CCFs) between the two light curves of the four band pairs ($u-g$, $u-r$, $u-i$, and $u-z$) were evaluated using an interpolation method, where the unevenly sampled light curves are linearly interpolated to calculate correlation coefficients at arbitrary lags (\cite{gas86})\footnote{We used {\tt PYCCF}~V2, Python Cross Correlation Function for reverberation mapping studies, for the cross-correlation analyses (\cite{sun18b}).}.
For each of the band pairs, a CCF was first calculated by interpolating the $u$-band light curve to the other band light curve sampling.
Then, another CCF was calculated by interpolating the latter light curve to the $u$-band light curve sampling.
Finally, a further CCF was calculated as an average of the two CCFs.
CCFs were calculated at 0.01-day intervals within the range of $-100 \leq \tau_{\rm obs}~{\rm [days]} \leq +100$, where a positive lag denotes a positive lag relative to the $u$-band light curve.
An estimate of the lag was evaluated as the centroid of the CCF.
The CCF centroid calculation was performed only for CCF points within 80\% of the peak value of the CCF.
Moreover, another estimate of the lag was evaluated as the peak of the CCF.
The uncertainties on the CCFs and their centroids and peaks were estimated using the Flux Randomization and Random Subsample Selection (FR/RSS) method developed by \citet{pet04}.
A total of 10,000 Monte Carlo FR/RSS realizations for each band pair produced a cross-correlation centroid distribution (CCCD) and a cross-correlation peak distribution (CCPD) for the lag relative to the $u$-band, and the distribution widths were used as estimates of the measurement uncertainties of the lags.

Figure~\ref{fig:ccfs} presents the CCFs, CCCDs, and CCPDs of the $u-g$, $u-r$, $u-i$, and $u-z$ band pair light curves of PG~2308+098.
The CCCDs and CCPDs are generally in agreement with each other, but the CCPDs tend to be more affected by the local peak of the CCFs.
Then, as in Section~\ref{javelin_twoband}, we used Equation~\ref{lag_relative_to_u} to calculate probability distributions of $\tau_{{\rm rest,~2500}}$ from the CCCDs and CCPDs of the $u-g$, $u-r$, $u-i$, and $u-z$ band pairs.
Figure~\ref{fig:all_lag} presents the probability distributions of $\tau_{{\rm rest,~2500}}$ derived from the CCCDs and CCPDs, along with the probability distributions derived from the JAVELIN thin disk and JAVELIN two-band analyses.
The median and $\pm 1\sigma$ values of the CCCD-based and CCPD-based probability distributions of $\tau_{{\rm rest,~2500}}$ are evaluated to be $\tau_{{\rm rest,~2500}} = 10.99^{+2.35}_{-2.52}$~days and $10.72^{+3.09}_{-3.19}$~days, and the 95th percentile upper limit is $14.84$~days and $15.74$~days, respectively (Table~\ref{disksize}).

\subsection{Comparison of JAVELIN thin disk, JAVELIN two-band, and CCCD/CCPD results}
\label{sec:comparison}

It should be noted that the widths of the CCCD/CCPD distributions shown in Figure~\ref{fig:ccfs} are larger than those of the posterior distributions of the lags derived by JAVELIN shown in Figure~\ref{fig:javelin_posterior}.
JAVELIN is known to generally yield smaller measurement uncertainties on the lag values compared to the CCF uncertainties evaluated with the FR/RSS method (e.g., \cite{fau16}), and this probably reflects the fact that the cross-correlation analysis does not assume any time series model for the two light curves.
Nevertheless, the trend that the lags become larger at longer wavelength bands, as seen in the results of the JAVELIN two-band analysis is also seen in the cross-correlation results.

\begin{figure}[tbp]
\center{
\includegraphics[clip, width=3.2in]{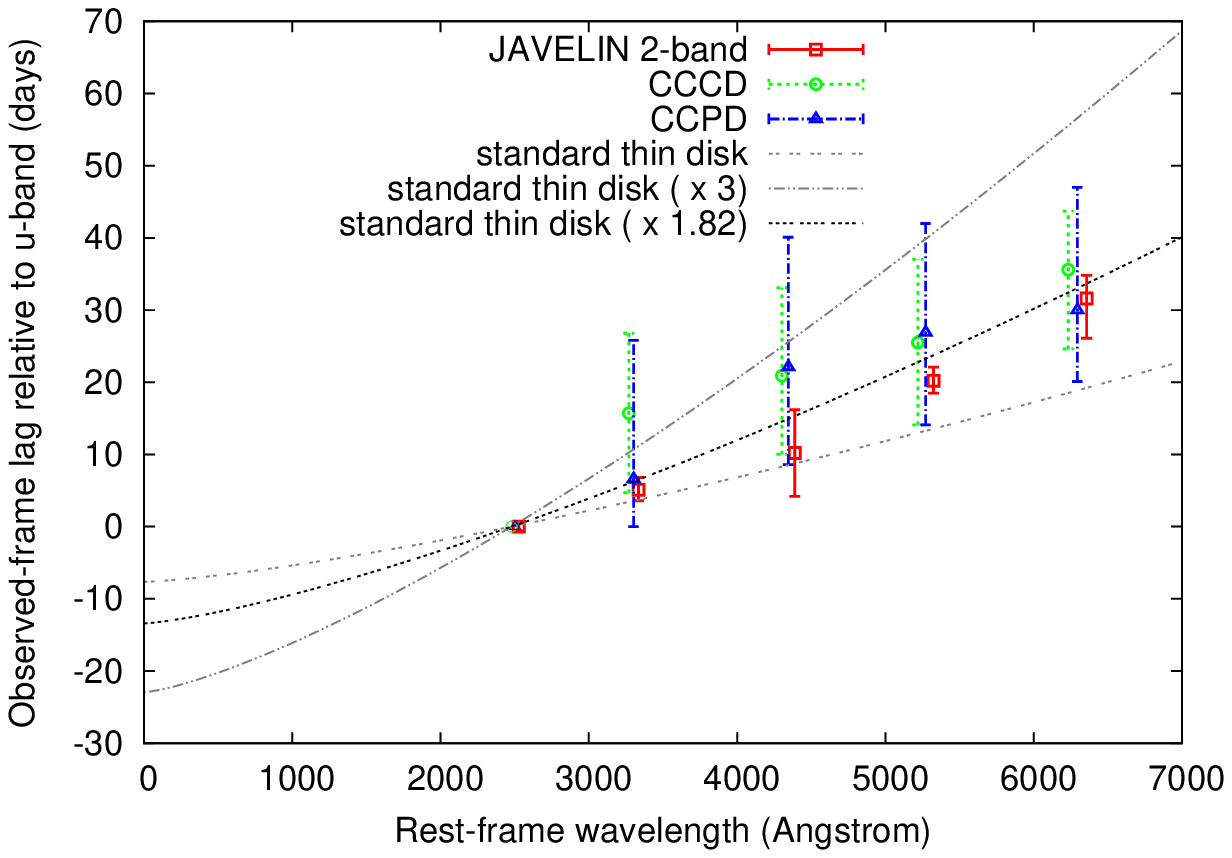}
}
 \caption{Comparisons of the continuum lags of PG~2308+098 measured with the JAVELIN two-band analysis and the CCF (CCCD and CCPD) analysis. All lags are measured relative to the $u$-band light curve ($\lambda_{u, {\rm rest}} = 2478$~\AA.).
  The observed-frame standard thin disk prediction (Equation~\ref{eqn:tau_rest} with $M_{\rm BH}=10^{9.6}M_{\odot}$ and $\dot{M}/\dot{M}_{Edd} = 10^{-0.9}$), and that inflated by a factor of $3$ and $1.82$, are also plotted (see Section~\ref{sec:disksize}).}
 \label{fig:lag_wave_4c0972}
\end{figure}

Figure~\ref{fig:lag_wave_4c0972} summarizes the lag estimates at $g$-, $r$-, $i$-, and $z$-bands relative to the $u$-band derived from the JAVELIN two-band and the CCF analyses for $u-g$, $u-r$, $u-i$, and $u-z$-band light curve pairs.
The lag spectra as a function of the rest-frame wavelength obtained from these two-band analyses are consistent with each other within the measurement uncertainties.
As noted in Section~\ref{sec:obs}, there may be a contribution from the hydrogen Balmer continuum emission to the lag spectra at the $g$-band.
However, any clear $g$-band excess is identified in the observed lag spectra, suggesting that the Balmer continuum contribution, if any, is negligible in the case of PG~2308+098.

\begin{figure}[tbp]
\center{
\includegraphics[clip, width=3.2in]{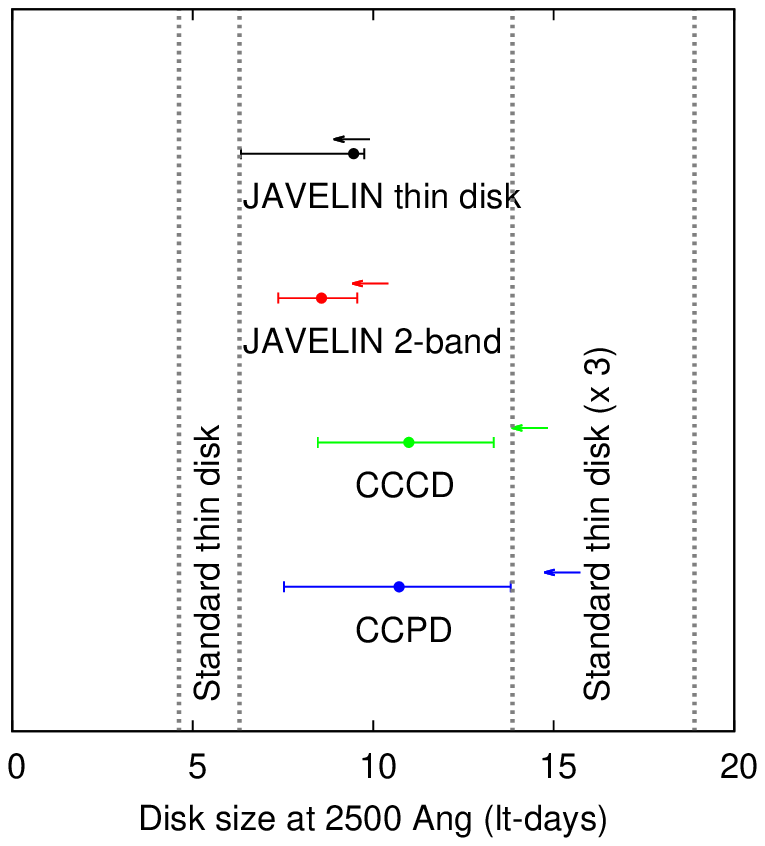}
}
 \caption{Observational constraints on the disk size at $\lambda_{\rm rest}=2500$~\AA\ for PG~2308+098 evaluated with different lag estimation methods (Figure~\ref{fig:all_lag}). 
 Data points denote the point estimates and 1$\sigma$ ranges, and left-facing arrows denote 95~\% percentile upper limits on the disk size.
 The standard thin disk model prediction range in the case of $M_{BH}=10^{9.6 \pm 0.1}$~$M_{BH}$ and $\dot{M}/\dot{M}_{Edd} = 10^{-0.9 \pm 0.1}$, and that inflated by a factor of $3$, are indicated by vertical lines.}
 \label{fig:comparisons}
\end{figure}

Figure~\ref{fig:comparisons} compares the constraints on the disk sizes at $\lambda_{\rm rest} = 2500$~\AA\ (i.e., $R_{\rm disk} = c \tau_{\rm rest, 2500}$) obtained from the results of the JAVELIN thin disk (Section~\ref{javelin_thin_disk}), JAVELIN two-band (Section~\ref{javelin_twoband}), CCPD and CCCD analyses (Section~\ref{sec:ccf}) on the assumption of $\beta=4/3$  (Table~\ref{disksize}).
Again, all disk size estimates are consistent with each other within the error bars.
As shown in Figure~\ref{fig:comparisons}, the 95~\% upper limits on the disk size estimates from these analyses strongly suggest that the disk size of PG~2308+098 is not unexpectedly large compared to the standard thin disk prediction.
Detailed comparisons among the observed disk size in PG~2308+098, the disk sizes in other reverberation-mapped or microlensing quasars, and the standard thin disk model will be presented in Section~\ref{sec:disksize}.

\section{Discussion}

\subsection{The disk size - BH mass relation}
\label{sec:disksize}

\begin{table*}
\tbl{The disk size of PG~2308+098 at $\lambda_{\rm rest}=2500$~\AA\ measured with several analysis methods.}{%
\begin{tabular}{lcccc}  
\hline\noalign{\vskip3pt} 
\multicolumn{1}{c}{Method} & $R_{\rm disk}$(2500~\AA) (lt-days) & 1$\sigma$ range (lt-days) & 2$\sigma$ range (lt-days) & 95th percentile upper limit (lt-days) \\  [2pt]
\hline\noalign{\vskip3pt}
JAVELIN thin disk   & 9.46  & 6.34---9.75  & 1.69---10.09 &  9.91 \\
JAVELIN 2-band lags & 8.57  & 7.37---9.56  & 6.09---11.10 & 10.42 \\
CCCD                & 10.99 & 8.47---13.34 & 5.74---15.69 & 14.84 \\
CCPD                & 10.72 & 7.53---13.81 & 4.08---16.87 & 15.74 \\ [2pt] 
\hline\noalign{\vskip3pt}
\end{tabular}}\label{disksize}
\begin{tabnote}
\hangindent6pt\noindent
\hbox to6pt{\footnotemark[$*$]\hss}\unskip%
$\pm 1\sigma$ and $\pm 2\sigma$ are defined as 68.2~\% and 95.4~\% ranges around a median of a probability distribution.
\end{tabnote}
\end{table*}

In the disk reprocessing picture, a variable X-ray/FUV emission from a compact central illumination source is irradiating the disk surface and continuously modifying the emissivity of the disk.
The lags of the reprocessed UV-optical light curves relative to the illuminating X-ray/FUV light curves correspond to the light-crossing times of the accretion disk.
Assuming the temperature profile of the standard thermal thin disk model, $T(R_{\rm disk})^4 = (3 GM_{\rm BH}\dot{M}/8\pi \sigma_{\rm SB}R_{\rm disk}^{3})$, and a lamppost geometry for the illuminating source, the disk radius responsible for the disk continuum emission at $\lambda_{\rm rest}$ can be expressed as:
\begin{eqnarray}
R_{\rm disk} &=& \left[ \left(\frac{hc}{k \lambda_{\rm rest}X}\right)^{-4} \left( \frac{3GM_{\rm BH}\dot{M}f_i}{8\pi \sigma_{\rm SB}} \right) \right]^{1/3} \nonumber \\
&=& 4.29~{\rm light~days} \nonumber \\
&\times& \left(\frac{\lambda_{\rm rest}}{2500~{\rm \AA}}\right)^{4/3} \times \left(\frac{M_{\rm BH}}{10^{9}~M_{\odot}}\right)^{2/3} \times \left(\frac{\dot{M}}{\dot{M}_{\rm Edd}}\right)^{1/3}\nonumber\\
&\times& \left(\frac{\eta}{0.1}\right)^{-1/3} \times \left(\frac{X}{2.49}\right)^{4/3} \times \left(\frac{f_i}{1}\right)^{1/3},
\label{eqn:tau_rest}
\end{eqnarray}
(e.g., \cite{bag95,cac07,mor10,fau16}) and the observed-frame lag relative to the illuminating light curve is $\tau_{\rm obs} = (1+z) \times \tau_{\rm rest} = (1+z) \times R_{\rm disk}/c $.
Here, $\sigma_{\rm SB}$ is the Stefan-Boltzmann constant, $\dot{M}$ is the mass accretion rate, and $\dot{M}_{\rm Edd}$ is the Eddington mass accretion rate.
The factor $X$ is the conversion factor from $T$ to $\lambda_{\rm rest}$ for a given $R_{\rm disk}$ [$T(R_{\rm disk}) = hc/k\lambda_{\rm rest}X$], which deviates from unity because a range of radii contribute to emission at $\lambda_{\rm rest}$ (\cite{col99,fau16}).
The exact value of $X$ depends slightly on the definition of the weighting \citep{col99,ede15,nod16,fau16,cac18}; here, we adopt $X=2.49$ following the definition given by \citet{mud17} (see Section~\ref{sec:implication} for further discussion).
With the introduction of the factor $X=2.49$, $R_{\rm disk}$ in Equation~\ref{eqn:tau_rest} represents the flux-weighted disk radius.
$\eta$ is the disk's radiative efficiency and can have a value between $0.04 \lesssim \eta \lesssim$ 0.3 for a BH spin Kerr parameter between $-1 < a < 0.998$ \citep{tho74}.
We choose $\eta=0.1$ as a fiducial value, which is adequate for a non-rotating Schwarzschild BH.
$f_i$ is defined as $f_i \equiv (3+\kappa)/3 \geq 1$, where $\kappa$ is a factor that represents the ratio of local viscous heating to external radiative heating from the X-ray/FUV irradiation.
Here, we adopt a value of $\kappa = 0$ ($f_i = 1$) (e.g., \cite{jia17}).
With this assumption, local viscous heating remains the dominant heating source of the disk, and the observationally-defined Eddington ratio $L_{\rm bol}/L_{\rm Edd}$ is directly comparable to the Eddington ratio defined via the mass accretion rate $\dot{M}/\dot{M}_{\rm Edd}$.
Although $\kappa = 1$ is usually assumed for local Seyfert galaxies (\cite{fau16,cac18}), the assumption of $\kappa = 0$ is considered to be valid for luminous quasars (e.g., \cite{mud17}).
The power-law dependences of $R_{\rm disk}$ on the parameters in Equation~\ref{eqn:tau_rest} are moderate, and thus the uncertainties related to the BH mass, the Eddington ratio, and the other model parameters do not severely affect the model predictions for $R_{\rm disk}$.

Figure~\ref{fig:radius_mass} plots thin disk model sizes at $\lambda_{\rm rest}=2500$~\AA\ (Equation~\ref{eqn:tau_rest}) for an Eddington ratio of $\dot{M}/\dot{M}_{\rm Edd} =$ 0.1, 0.3, and 1.0.
At a BH mass of $M_{BH}=10^{9.6 \pm 0.1} M_{\odot}$ and an Eddington ratio of $\dot{M}/\dot{M}_{\rm Edd} = 10^{-0.9 \pm 0.1} = 0.13 \pm 0.03$ estimated for PG~2308+098 (Section~\ref{object}), the disk size at $\lambda_{\rm rest} = 2500$~\AA\ is expected to be 
\begin{equation}
R_{\rm disk}(2500~{\rm \AA}) = 5.46~(\pm 0.84)~{\rm light\mathchar`-days},
\end{equation}
based on the standard thin disk model.
In contrast, the median and 1$\sigma$ range of the disk size at $\lambda_{\rm rest} = 2500$~\AA\ of PG~2308+098 constrained from the JAVELIN thin disk model are $9.46$ and $6.34-9.75$ light-days, respectively (Table~\ref{disksize}), which correspond to an Eddington ratio of $\dot{M}/\dot{M}_{\rm Edd} = 0.68$ and $0.20 - 0.74$, respectively.
Moreover, the 95th percentile upper limit for the disk size at $\lambda_{\rm rest} = 2500$~\AA\ of PG~2308+098 is 9.91 light-days, corresponding to an Eddington ratio of $\dot{M}/\dot{M}_{\rm Edd} = 0.78$.
Therefore, the disk size of PG~2308+098 as measured by continuum reverberation mapping is slightly larger than the standard thin disk prediction (the 95~\% upper limit of the size ratio is $9.91/5.46=1.82$), and remains consistent with the sub-Eddington accretion rate standard thin disk model.

Figure~\ref{fig:radius_mass} compares the continuum reverberation measurements of the accretion disk sizes at $\lambda_{\rm rest}=2500$~\AA\ of PG~2308+098 and those derived by \citet{mud17} for the 15 DES quasars\footnote{7 quasars have lag detections in multiple years, thus there are 24 data points.} as a function of the BH masses.
The disk size measurements of \citet{mud17} are taken directly from Table~3 in \citet{mud17}, where the associated uncertainties are $\pm 2 \sigma$ ranges (also \cite{mud17b}).
Note that the disk size constraints for the 15 DES quasars by \citet{mud17} and for PG~2308+098 obtained in this work and shown in Figure~\ref{fig:radius_mass} are derived using the same analysis technique (the JAVELIN thin disk model fitting; Section~\ref{javelin_thin_disk}); thus, they are directly comparable to each other.
Additionally, the $g-r$, $g-i$, and $g-z$ two-band lags of the $39$ Pan-STARRS quasars (derived using JAVELIN) listed in Table~3 of \citet{jia17} were converted into the rest-frame 2500~\AA\ disk size by applying the same procedure described in Section~\ref{javelin_twoband} (assuming $\beta=4/3$) and plotted in Figure~\ref{fig:radius_mass}.
It is important to note that the optical continuum reverberation constraints on the disk sizes at $\lambda_{\rm rest} = 2500$~\AA\ (\cite{jia16,mud17}, and this work) are directly affected by the choice of $\beta$ due to the $\beta$ dependence of the extrapolation of the lag spectrum to driving wavelengths ($\lambda \rightarrow 0$~\AA).

Figure~\ref{fig:radius_mass} also presents the accretion disk sizes at $\lambda_{\rm rest}=2500$~\AA\ of gravitationally-lensed quasars reported in \citet{mor10} and derived from microlensing analysis.
The analysis in \citet{mor10} is based on time-series observations of microlensing variability and sets constraints on the scale length $R_{s}$ of the accretion disks.
Note that the disk scale length $R_{s}$ is equivalent to the case of $X=1$ in Equation~\ref{eqn:tau_rest} (see \cite{mor10}).
Although the reverberation-mapped disk radii are insensitive to disk inclination angles (e.g., \cite{starkey17,hal18}), the microlensing disk size measurements directly depend on the inclination angle as $\propto \sqrt{\cos i}$, and therefore the disk sizes presented in \citet{mor10} were corrected for a mean inclination $<\cos i> = 1/2$ to obtain face-on estimates.
The disk sizes presented in \citet{mor10} were corrected to $\lambda_{\rm rest}=2500$~\AA\ assuming $\beta=4/3$.
With regard to the measurements made by \citet{mor10}, the choice of $\beta$ had little effect on the disk size estimates at $\lambda_{\rm rest}=2500$~\AA\ because their microlensing observations were carried out at around the rest-frame wavelengths of $\lambda_{\rm rest} \simeq 2500$~\AA\ (see \cite{mor10} for details).
Then, to compare the microlensing-based disk scale lengths $R_{s}$ with the reverberation disk sizes, we inflated $R_{s}$ by a factor of $X^{4/3} = 2.49^{4/3}=3.37$, following Equation~\ref{eqn:tau_rest}, as in \citet{mud17}.

Figure~\ref{fig:radius_mass} clearly shows that the disk sizes for $M_{BH} \lesssim 5 \times 10^{8}~M_{\odot}$ quasars derived from continuum reverberation mapping and gravitational microlensing are both generally distributed above the standard thin disk predictions with sub-Eddington accretion rates, requiring super-Eddington accretion rates (\cite{hal14}).
On the other hand, a proportion of quasars with higher BH masses, including PG~2308+070 analyzed in this work, are consistent with sub-Eddington accretion rates.
\footnote{
Among the data from \citet{jia17} given in Figure~\ref{fig:radius_mass}, we can readily identify one very massive outlier quasar, SDSS~J083841.70+430519.0 ($M_{\rm BH} = 10^{10} M_{\odot}$ at $z=0.98$; \cite{jia17,koz17b}), which has a disk size a factor of 3 smaller compared to the standard thin disk model. This quasar is ignored in Figures~10 and 11 of \citet{mud17}. The rest-frame $g-r$, $g-i$, and $g-z$ two-band lags of SDSS~J083841.70+430519.0 are clearly wavelength-dependent ($1.09 \pm 0.25$, $1.59 \pm 0.25$, and $5.91 \pm 0.25$~days, respectively) and the disk size at 2500~\AA\ is robustly determined as $3.93 \pm 0.17$~light-days on the assumption of a value of $\beta=4/3$. However, the estimated disk radius is unphysically small considering the ISCO of a Schwarzschild BH of $M_{\rm BH} = 10^{10} M_{\odot}$, $R_{\rm ISCO} = 6GM_{\rm BH}/c^2 = 3.42$~light-days, suggesting that the BH mass estimate for this object is erroneous.\label{ftn:outlier}
}
Overall, the measured disk sizes appear to have a weaker dependence on $M_{\rm BH}$ than the expectations from the standard thin disk model ($R_{\rm disk} \propto M_{\rm BH}^{2/3}$ for a given Eddington ratio); this trend has already been noted by \citet{mud17} and \citet{bla11} in the context of continuum reverberation mapping and microlensing measurements, respectively.
\citet{bla11} pointed out that the mass dependence of the disk size at 2500~\AA\ is about half the expected dependence, i.e., $\propto M_{\rm BH}^{1/3}$, and this argument still holds for the combined dataset shown in Figure~\ref{fig:radius_mass}.

\subsection{Implications for the accretion disk structure in PG~2308+098}
\label{sec:implication}

\begin{figure*}[tbp]
\center{
\includegraphics[clip, width=6.2in]{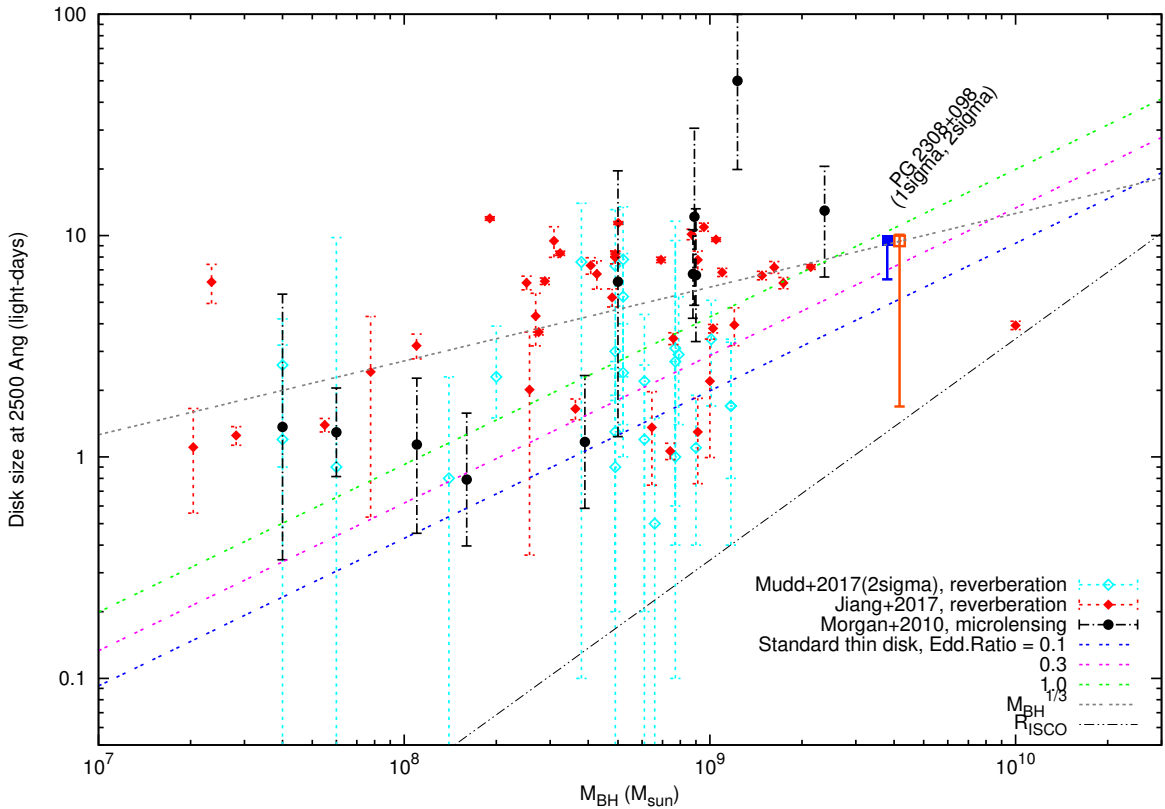}
}
 \caption{Flux-weighted disk radius at $\lambda_{\rm rest}=2500$~\AA\ vs. the black hole mass relation for AGNs/quasars. Filled and open squares denote, respectively, the $\pm 1\sigma$ and $\pm 2 \sigma$ disk size constraints for PG~2308+098 derived from the JAVELIN thin disk analysis on the assumption of $\beta=4/3$, and the two points are vertically shifted for clarity. The disk size constraints based on the continuum reverberation lags for quasar samples of \citet{jia17} and \citet{mud17}, and those based on the microlensing measurements taken from \citet{mor10}, are also plotted for comparison (see text in Section~\ref{sec:disksize} for details). The error-bars are $\pm 1\sigma$, except for the data points from \citet{mud17} where the error bars denote $\pm 2\sigma$ ranges. Note that $M_{\rm BH}$ (taken from each paper) generally have an uncertainty of 0.4~dex. The standard thin disk model prediction for Eddington ratios of 0.1, 0.3, and 1.0 are denoted by lines ($R_{\rm disk} \propto M_{\rm BH}^{2/3}$; Equation~\ref{eqn:tau_rest}), along with an arbitrarily scaled line of $R_{\rm disk} \propto M_{\rm BH}^{1/3}$, and the innermost stable circular orbit (ISCO; $R_{\rm ISCO} = 6GM_{\rm BH}/c^2$), for comparison. Note that the black hole mass estimate for the most massive quasar in the sample provided by \citet{jia17}, SDSS~J083841.70+430519.0 with $M_{\rm BH} = 10^{10}$~$M_{\odot}$, may be erroneous (see footnote~\ref{ftn:outlier}).}
 \label{fig:radius_mass}
\end{figure*}

Although we show that the discrepancy between the measured disk size of PG~2308+098 and the expected size from the standard disk theory is modest, it remains a general trend that the AGN disk sizes (probed either by microlensing events or continuum reverberation mapping) are generally larger than the standard disk predictions.
It should be noted that the presence of non-zero BH spins makes the disk size problem even more significant.
The `Soltan argument' (\cite{sol82,yu02,ued14}) suggests that modest radiative efficiencies of $\eta \gtrsim 0.1$ are required for accretion disks in a general population of luminous quasars to account for BH mass density in a low-redshift universe.
The radiative efficiency, $\eta$, is an increasing function of the BH spin, so the Soltan argument may imply that the quasars generally harbor prograde spinning SMBHs (\cite{yu02}).
If we consider prograde spinning SMBHs ($a>0$), the difference between the observations and the model becomes even more significant because a prograde BH accretion disk has a higher radiative efficiency ($\eta > 0.1$) which leads to a smaller accretion disk compared to a non-spinning BH disk for a given luminosity or Eddington ratio (see Equation~\ref{eqn:tau_rest}).
A maximally prograde SMBH has $\eta \simeq 0.3$, and the accretion disk size is expected to be smaller than the non-spinning SMBH accretion disk by a factor of $(0.3/0.1)^{-1/3} = 0.69$.
On the other hand, a maximally retrograde SMBH has $\eta \simeq 0.04$, and the disk size is larger by a factor of $(0.04/0.1)^{-1/3} = 1.36$, but there is no physical motivation to consider retrograde SMBHs in actual quasars.

It is important to note that the power-law index for $M_{\rm BH}$ and $\dot{M}/\dot{M}_{\rm Edd}$ in Equation~\ref{eqn:tau_rest} is so small that the disk size discrepancy requires huge modifications of the assumed parameters. 
For example, to inflate the theoretical disk size by a factor of $3$ for a given $M_{\rm BH}$, a value of the Eddington ratio $3^{3}=27$ times larger is required, which easily violates the basic assumption that normal AGNs/quasars have sub-Eddington accretion rates and can be described by the standard thin disk theory.
As for the model parameters, the choice of the factor $X$ in Equation~\ref{eqn:tau_rest} potentially has non-negligible effects on the standard disk predictions for $R_{\rm disk}$ (see \cite{cac18} for a related discussion).
Interestingly, if we adopt $X=3.89$ as suggested by \citet{col99}, which is based on detailed calculations of the disk response to continuum emission, the expected disk size increases by a factor of $(3.89/2.49)^{4/3} = 1.81$, and therefore the measured disk size of PG~2308+098 is in complete agreement with the standard disk prediction.
Nevertheless, the disk sizes in other less massive AGNs/quasars shown in Figure~\ref{fig:radius_mass} remain larger than the standard disk prediction, even when adopting a value of $X=3.89$.
Therefore, the larger-than-expected accretion disk sizes, and the shallower slope of the disk size vs. $M_{\rm BH}$ relation compared to the standard thin disk theory confirmed by our disk size measurement for PG~2308+098, may suggest that part of our basic assumption based on the standard thin disk theory is invalid for describing quasar accretion disks.

The power-law index $\beta=4/3$ (Equation~\ref{eqn:temp_radius}) assumed throughout this work is taken directly from the standard thin disk model prediction of the disk temperature profile of $T(R) \propto R^{-3/4}$.
As noted in Section~\ref{object}, the spectral shape of the polarized flux component and the variable component of PG~2308+098 at the rest-frame UV-optical wavelengths suggests that the assumption of $\beta=4/3$ is not far from reality.
Nevertheless, as discussed in Section~\ref{javelin_thin_disk}, the multi-band optical continuum reverberation mapping observation alone cannot strongly constrain $\beta$ (see also \cite{mud17}).
Therefore, it may still be possible that the wavelength dependence index $\beta$ can differ from a value of $4/3$.
Interestingly, \citet{jia17} found tentative evidence of $\beta < 4/3$, particularly for the high-luminosity quasars they analyzed.
Moreover, \citet{bla11} noted that microlensing constraints on $\beta$ for quasars on average imply a value of $\beta < 4/3$ (see also \cite{hal14}).
If we adopt a value of $\beta$ smaller than $4/3$, the rest-frame 2500~\AA~disk size estimates from the continuum reverberation measurements become generally larger than those shown in Figure~\ref{fig:radius_mass}.
However, as noted in Section~\ref{sec:disksize}, the microlensing constraints on the rest-frame 2500~\AA~disk sizes are, in principle, largely independent of $\beta$ (see also \cite{mor05}).
Therefore, $\beta < 4/3$ makes the disk size estimates from the continuum reverberation and microlensing measurements inconsistent with each other.
Currently, such a discrepancy is not obviously observed in Figure~\ref{fig:radius_mass}.
Moreover, smaller $\beta$ results in a steeper temperature radial profile and a bluer UV-optical spectrum, which may contradict with the observed AGN/quasar UV-optical spectra (e.g., \cite{gas08}).
Further continuum reverberation/microlensing observations are needed to determine the true value of $\beta$.

Another potentially important point of view concerns the metallicity of accretion disks in AGNs.
\citet{jia17} claimed that quasars with higher metallicity (probed as the metallicity of broad line regions) appear to have systematically smaller continuum reverberation lags, i.e., smaller disk sizes.
The relationship between disk size and metallicity may possibly be explained by large changes in the disk's opacity as a function of the gas metallicity due to the iron opacity bump, which can significantly modify the thermal properties and structure of AGN accretion disks compared to the standard thin disk model (\cite{jia16,jia17}).
Because the higher BH masses generally imply a higher broad line region gas metallicity (\cite{mat11} and references therein), the above claim may provide another perspective on the weaker $M_{\rm BH}$ dependence of $R_{\rm disk}$ than the standard disk model prediction discussed in Section~\ref{sec:disksize}.
According to this interpretation, the relatively smaller disk size observed in PG~2308+098 implies an intrinsically different disk structure in this high BH mass quasar compared to lower BH mass quasars.
As suggested by \citet{jia17}, the disk size problem revealed by the continuum reverberation and microlensing observations may indicate a need for modifications to the standard disk theory so that metallicity-dependent opacity effects are appropriately included.

Finally, it should be noted that the original idea of continuum reverberation mapping is to measure the time lag between the driving X-ray light curve and the responding UV-optical light curves, on the assumption of the reprocessing picture where the X-ray/FUV emission from the central compact X-ray/FUV emission region (`lamppost') is irradiating the accretion disk surface.
Continuum reverberation mapping observations for local well-studied Seyfert galaxies, e.g., NGC~4395 (\cite{mch16}), NGC~4593 (\cite{pal18},\cite{cac18}), NGC~5548 (\cite{ede15,fau16}), and NGC~4151 (\cite{ede17}), were based on this idea and measured the X-ray-to-UV/optical lag as a direct probe of the disk sizes.
These X-ray-to-UV/optical continuum reverberation mapping observations revealed that the disk sizes are several times larger than the standard thin disk prediction (see also \cite{bui17}).
On the other hand, the continuum reverberation disk size constraints discussed in this work (\cite{jia17,mud17}, and this work for PG~2308+098) are based on wavelength-dependent reverberation lag measurements within the rest-frame UV-optical bands.
\citet{mch17} pointed out that although X-ray-to-UV/optical reverberation lags are observed to be several times larger in local Seyfert AGNs, the wavelength-dependent continuum lags of these AGNs within the UV-optical bands are generally more consistent with the reprocessing picture utilizing the standard thin disk model (see also \cite{ede17}).
In other words, the X-ray light curves of AGNs alone appear to show extra offset lags relative to simple disk reprocessing.
\citet{mch17} argued that the extra offset lag in the X-ray band might possibly be explained by introducing an additional distant X-ray reflector, probably broad emission line region clouds, which can dilute the X-ray-to-UV/optical disk reprocessing signals.
Alternatively, the X-ray emission might first be reprocessed into soft X-ray/FUV photons at around the innermost disk region, and then the FUV photons might be reprocessed into UV-optical disk photons (\cite{gar17,ede17}).
Moreover, truncation of the inner accretion disks may produce an additional X-ray-to-UV/optical lag relative to the lamppost disk reprocessing model prediction \citep{nod16}.
The wavelength-dependent UV-optical inter-band lag measurements are free from such uncertainties regarding the geometry of the X-ray/FUV emitter; nevertheless, as shown in Figure~\ref{fig:radius_mass}, the accretion disk sizes inferred from the UV-optical inter-band lag measurements remain larger than the standard thin disk predictions.

The larger disk sizes in each of the AGNs, and the shallower slope of the global $R_{\rm disk}-M_{\rm BH}$ relation compared to the standard thin disk model, suggest that this model cannot adequately describe AGN/quasar accretion disks.
However, we wish to point out that the compilation of data from the literature and the discussion in Section~\ref{sec:cont_rev_measure} suggest that differences between analysis methods used may introduce systematic biases into the disk size constraints (see also \cite{bla11} for the case of microlensing measurements).
Homogeneous re-analysis of data from the literature will be useful to further investigate the possible systematics of the disk size estimations.
Moreover, future X-ray-to-UV/optical and UV-optical inter-band continuum reverberation mapping observations, as well as microlensing observations, for a statistical sample of AGNs/quasars with high cadence, high quality light curve data are needed to obtain conclusive results on the (in)validity of the standard disk theory and to search for a solution to the disk size problem.
The Large Synoptic Survey Telescope (LSST) will automatically obtain high-cadence multi-band optical light curves for millions of quasars, including thousands of gravitationally lensed quasars; thus, LSST will enable optical continuum reverberation measurements and microlensing measurements for a large sample of quasars (e.g., \cite{lsst09,ogu10}).

\section{Conclusions}
\label{sec:conclusion}

We carried out $u$-, $g$-, $r$-, $i$-, and $z$-band monitoring observations for the massive quasar PG~2308+098 using the Kiso Schmidt telescope/KWFC and detected wavelength-dependent continuum reverberation lags at $g$-, $r$-, $i$-, and $z$-bands relative to the $u$-band, where the longer wavelength bands lag behind the shorter wavelength bands.
On the assumption of the lamppost X-ray/FUV reprocessing picture and the standard thin disk temperature profile of $T(r) \propto R^{-3/4}$ (i.e., $R_{\rm disk} = c\tau_{\rm rest}  \propto \lambda^{4/3}$), we derived constraints on the flux-weighted radius of the accretion disk at $\lambda_{\rm rest} = 2500$~\AA, as $R_{\rm disk} = 6.34-9.75$~light-days ($1\sigma$ range from the JAVELIN thin disk analysis).
We confirmed that the disk size estimates from the JAVELIN thin disk, JAVELIN two-band, and two-band CCF (CCCD/CCPD) analyses are consistent with each other within the errors.

The accretion disk sizes of AGNs measured via the continuum reverberation and microlensing techniques are known to be generally $\sim 3$ times larger than the standard disk model predictions (e.g., \cite{mor10,ede15,fau18}).
For PG~2308+098, the measured disk size is also larger, although modestly, by a factor of $1.2-1.8$ times that of the standard disk model prediction.
The compilation of literature data, including the new disk size constraint for PG~2308+098, revealed that the global $R_{\rm disk}$(2500\AA)-$M_{\rm BH}$ relation of AGNs/quasars may have a shallower slope compared to the standard accretion disk model prediction of $R_{\rm disk} \propto M_{\rm BH}^{2/3}$.
The larger disk sizes and the shallower $R_{\rm disk}$(2500\AA)-$M_{\rm BH}$ slope imply that the standard accretion disk model cannot properly account for AGN/quasar accretion disks. 
One potentially important factor to be considered is the metallicity of the accretion disks (e.g., \cite{jia16,jia17}).
Construction of a sophisticated accretion disk theory with appropriate treatment of the opacity effects (including the iron opacity jump), as well as further observations of the continuum reverberation and microlensing events for AGN/quasar accretion disks, will be needed to tackle the disk size problem.

\begin{ack}

We thank Yuki Sarugaku for his support during the KWFC queue mode observations.
We are grateful to all the staff of the Kiso Observatory for their efforts to maintain the observation system.
This work was supported by JSPS KAKENHI Grant Number 17J01884.

This research has made use of NASA's Astrophysics Data System Bibliographic Services.
This research has made use of the NASA/IPAC Extragalactic Database (NED), which is operated by the Jet Propulsion Laboratory, California Institute of Technology, under contract with the National Aeronautics and Space Administration.

This research made use of Astropy, a community-developed core Python package for Astronomy \citep{ast13}.

\end{ack}

\bibliography{qpv}

\end{document}